\documentclass[12pt]{article}
\def\bea{\begin{eqnarray}}
\def\ve{\vert}
\def\eea{\end{eqnarray}}

\def\nnb{\nonumber}
\def\la{\langle}
\def\ra{\rangle}
\def\ba{\begin{array}}
\def\ea{\end{array}}
\def\tr{\mbox{Tr}}
\begin{document}
\title
{\bf Magnetic Moments of $\Delta$ Baryons in Light Cone QCD Sum Rules}
\author{T.M. Aliev, A. {\"O}zpineci, M. Savc{\i}} 
\begin{titlepage}
\maketitle
\begin{abstract}

We calculate the magnetic moments of $\Delta$  baryons within the framework of QCD sum rules.
A comparison of our results on the magnetic moments of the $\Delta$ baryons with the predictions of
different approaches is presented.
\end{abstract} 
\end{titlepage}

\section{Introduction}
The extraction of the fundamental parameters of hadrons from experimental data requires some information
about physics at large distances and they can not be calculated directly from fundamental QCD
Lagrangian because at large distance strong coupling constant, $\alpha_s$, becomes large and
perturbation theory is invalid.  For this reason for determination of hadron parameters, a reliable
non-perturbative approach is needed.  Among other non-perturbative approaches, QCD sum rules
\cite{shifman} is an especially powerful method in studying the properties of low-lying hadrons.  In
this method, deep connection between the hadron parameters and the QCD vacuum structure is established
via a few condensate parameters.  This method is adopted and extended in many works (see for example
Refs. \cite{reinders, ioffe1, chung} and references therein). One of characteristic parameters of the
hadrons is their magnetic moments. Calculation of the nucleon magnetic
moments in the framework of QCD sum rules method using external fields technique, first suggested in
\cite{ioffe2}, was carried out in \cite{ioffe3, balitsky1}.  They were later refined and extended to the
entire baryon octet in \cite{chiu, pasupathy}.

Magnetic moments of the decuplet baryons are calculated in \cite{belyaev2,lee1} within the framework of
QCD sum rules using external field.  Note that in \cite{belyaev2}, from the decuplet baryons, only the
magnetic moments of $\Delta^{++}$ and $\Omega^-$ were calculated.  At present, the magnetic moments of
$\Delta^{++}$ \cite{bosshard}, $\Delta^{0}$ \cite{heller} and $\Omega^-$
\cite{wallace} are known from
experiments.  The experimental information provides new incentives for theoretical  scrutiny
of these physical quantities.

In this letter, we present an independent calculation of the magnetic moments of $\Delta^{++}$,
$\Delta^{+}$, $\Delta^{0}$, and $\Delta^{-}$ within the framework of an alternative approach to the
traditional sum rules, i.e. the light cone QCD sum rules (LCQSR).
Comparison of the  predictions of this approach on magnetic moments with the results of
other methods existing in the literature, and the experimental results is also presented.

The LCQSR is based on the operator product expansion on the light cone, which is an expansion over the
twists of the operators rather than dimensions as in the traditional QCD sum rules.  The main
contribution comes from the lower twist operator.  The matrix elements of the nonlocal operators between
the vacuum and hadronic state defines the hadronic wave functions. (More about this method and its
applications can be found in \cite{braun2, braun22} and references therein).  Note that magnetic moments
of the nucleon using LCQSR approach was studied in \cite{braun3}.

The paper is organized as follows.  In Sect. II, the light cone QCD sum rules for the magnetic moments
of $\Delta^{++}$, $\Delta^+$, $\Delta^0$, and $\Delta^-$ are derived.  In Sect. III, we present our
numerical analysis and conclusion.

\section{Sum Rules for the Magnetic Moments of $\Delta$ baryons}
According to the QCD sum rules philosophy, a quantitative estimate for the $\Delta$ magnetic moment can
be obtained by equating two different representations of the corresponding correlator, written in terms
of hadrons and quark-gluons. For this aim, we consider the following correlation  function

\bea
\Pi_{\mu \nu} = i \int dx e^{i p x} \la 0 \ve {\cal T} \eta^B_\mu(x) \bar \eta^B_\nu(0) \ve 0
\ra_\gamma
\label{correlation}\, ,
\eea
where ${\cal T}$ is the time ordering operator, $\gamma$ means external electromagnetic field.
In this expression  the $\eta^B_\mu$'s are the interpolating currents for the baryon B. 
This correlation function can be calculated on one side phenomenologically, in terms of the hadron
properties and on the other side by the operator product expansion (OPE) in the deep Eucledian
region of the correlator momentum $p^2 \rightarrow - \infty$ using QCD degrees of freedom.
By equating both expressions, we construct the corresponding sum rules.

On the phenomenological side, by inserting a complete set of one hadron states into the correlation
function, Eq. (\ref{correlation}), one obtains:
\bea
\Pi_{\mu \nu}(p_1^2,p_2^2) = \sum_{B_1, \, B_2} \frac{\la 0 \ve \eta^B_{\mu} \ve
B_1(p_1) \ra}{p_1^2 - M_1^2} 
\la B_1 (p_1) \ve  B_2(p_2) \ra_\gamma 
\frac{\la B_2 (p_2) \ve \eta^B_\nu \ve 0 \ra}{p_2^2 - M_2^2},
\label{insert}
\eea 
where $p_2 = p_1 + q$, $q$  is the photon momentum, $B_i$ form a complete set of baryons having the same
quantum numbers as $B$, with masses $M_i$.  

The matrix elements of the interpolating currents between the ground state and
the state containing a single baryon, $B$, with momentum $p$ and having spin $s$ is defined as:
\bea
\la 0 \ve \eta_{\mu} \ve B(p,s) \ra = \lambda_B u_\mu (p,s) \, , \label{3}
\eea
where $\lambda_B$ is a phenomenological constant parametrizing the coupling strength of the baryon to
the current, and $u_\mu$ is the Rarita-Schwinger spin-vector satisfying $(\not\! p -M_B) u_\mu = 0$,
$\gamma_\mu u_\mu = p_\mu u_\mu = 0$. (For a discussion of the properties of the Rarita-Schwinger
spin-vector see e.g. \cite{taka}).
In order to write down the phenomenological part of the sum rules, one also needs an expression for the
matrix element $\la B(p_1) \ve B (p_2) \ra_\gamma$.  In the general case, the electromagnetic vertex of
spin $3/2$ baryons can be written as
\bea
\la B(p_1) \ve B(p_2) \ra_\gamma = \epsilon_\rho \bar u_\mu(p_1) {\cal O}^{\mu \rho
\nu}(p_1,p_2) u_\nu (p_2) \, , \label{4}
\eea
where $\epsilon_\rho$ is the polarization vector of the photon and the Lorentz tensor ${\cal O}^{\mu
\rho \nu}$ is given by:
\bea
{\cal O}^{\mu \rho \nu}(p_1,p_2) &=& - g^{\mu \nu} \left[ \gamma_\rho (f_1+f_2) +
\frac{(p_1+p_2)_\rho}{2 M_B}
f_2 +q_\rho f_3 \right]  - \nnb \\
&-& \frac{q_\mu q_\nu}{(2 M_B)^2} \left[ \gamma_\rho (g_1+g_2) + \frac{(p_1 + p_2)_\rho}{2 M_B} g_2 +
q_\rho
g_3 \right] \label{5}
\eea
where the form factors $f_i$ and $g_i$ are (in the general case) functions of $q^2=(p_1 - p_2)^2$.  In
our problem, the values of the formfactors only at one point, $q^2=0$, are needed.

In our calculation, we also performed summation over spins of the Rarita-Schwinger spin vector,
\bea
\sum_s u_\sigma(p,s) \bar u_\tau(p,s) = - \frac{(\not\! p + M_B)}{2 M_B} \left\{ 
g_{\sigma \tau} - \frac{1}{3} \gamma_\sigma \gamma_\tau - \frac{2 p_\sigma p_\tau}{3 M_B^2} +
\frac{p_\sigma \gamma_\tau - p_\tau \gamma_\sigma}{3 M_B} \right\} \label{6}
\eea
Using Eqs. (\ref{3}-\ref{6}),
the correlation function can be expressed as the sum of various structures, not all of them
independent.  To remove the dependencies, an ordering of the gamma matrices should be chosen.   
For this purpose the structure $\gamma_\mu\!\!\not\!p_1\!\!\not\!\epsilon\!\!\not\!p_2 \gamma_\nu$ 
is chosen.  With this ordering, the correlation function becomes:
\bea
\Pi_{\mu \nu} &=& \lambda_B^2 \frac{1}{(p_1^2 - M_B^2)(p_2^2 - M_B^2)} \left[ g_{\mu \nu} 
\not\!p_1\!\not\!\epsilon\!\not\!p_2
\frac{g_M}{3} + \right. \nnb \\
&+&  \mbox{\it other structures with $\gamma_{\mu}$ at the beginning and  $\gamma_{\nu}$ at the
end \huge $]$}
\eea 
where $g_M$ is the magnetic form factor, $g_M/3 = f_1 +f_2$.  $g_M$ evaluated at $q^2 =0$ gives the
magnetic moment of the baryon in units of its natural magneton, $e \hbar/2 m_B c$.
The appearance of the factor $3$ can be understood from the fact that
in the nonrelativistic limit, the maximum energy of the baryon in the presence of a uniform magnetic
field with magnitude $H$  is $3(f_1 + f_2) H \equiv g_M \, H$ \cite{belyaev1}.  

The reason for choosing this structure can be explained as follows.
In general the interpolation current might also have a non-zero 
overlap with spin $\frac{1}{2}$ baryons, but spin $\frac{1}{2}$ baryons do not contribute to the
structure  
$g_{\mu \nu}\! \not\!p_1\!\!\not\!\epsilon\not\!p_2$ since their overlap is given by:
\bea
\la 0 \ve \eta_\mu \ve J=1/2 \ra = (A p_\mu + B \gamma_\mu) u(p)
\eea
where
$(\not\!p - m) u(p) = 0$ and $(A m + 4 B) = 0$ \cite{belyaev1,leinweber1}.

In order to calculate the correlator (\ref{correlation}) from the QCD side, first,
appropriate
currents should be chosen.  For the case of the
$\Delta$ baryons, they can be chosen as (see for example \cite{lee1}): 
\bea
\eta_\mu^{\Delta^{++}} &=& \epsilon^{abc} (u^{aT} C \gamma_\alpha u^b) u^c \, ,\nnb \\
\eta_\mu^{\Delta^{+}} &=& \frac{1}{\sqrt{3}} \epsilon^{abc} [2 (u^{aT} C \gamma_\alpha d^b) u^c +
(u^{aT} C \gamma_\alpha u^b) d^c] \, , \nnb \\ 
\eta_\mu^{\Delta^{0}} &=& \frac{1}{\sqrt{3}} \epsilon^{abc} [2 (d^{aT} C \gamma_\alpha u^b) d^c +
(d^{aT} C \gamma_\alpha d^b) u^c] \, , \nnb \\
\eta_\mu^{\Delta^{-}} &=& \epsilon^{abc} (u^{aT} C \gamma_\alpha u^b) u^c \, ,
\eea
where $C$ is the charge conjugation operator, $a, \, b,\, c$ are color indices.
It should be noted that these baryon currents are not unique, one can choose  an
infinite number of currents with the $\Delta$ baryon quantum numbers \cite{chung, ioffe4}.

On the QCD side, for the same correlation functions we obtain:
\bea
\Pi_{\mu \nu}^{\Delta^{++}} =&& \Pi_{\mu \nu}'^{\Delta^{++}}
+ \frac{1}{2} \epsilon^{abc} \epsilon^{def} \int d^4 x e^{ipx}\la \gamma(q) \ve \bar u^f A_i u^a \nnb \\
&&\left\{
2 S_u^{cd} \gamma_\nu S_u'^{be} \gamma_\mu A_i + 
2 S_u^{cd} \gamma_\nu A'_i \gamma_\mu S_u^{be} + 
\right. \nnb \\
&& \left.
+ 2 A_i \gamma_\nu S_u'^{cd} \gamma_\mu S_u^{be} +
S_u^{cd} \tr (\gamma_\nu {S'_u}^{be} \gamma_\mu A_i) +
\right. \nnb \\
&&\left.
+S_u^{cd} \tr (\gamma_\nu  A'_i \gamma_\mu S_u^{be}) +
A_i \tr (\gamma_\nu {S'_u}^{cd} \gamma_\mu S_u^{be})
\right\} \ve 0 \ra \label{deltapp} \\
\Pi_{\mu \nu}^{\Delta^{+}} =&& \Pi_{\mu \nu}'^{\Delta^{+}}
- \frac{1}{6} \epsilon^{abc} \epsilon^{def} \int d^4 x e^{ipx}\la \gamma(q) \ve \bar u^d A_i u^a \nnb \\
&&\left\{
2 A_i \gamma_\nu {S'_d}^{be} \gamma_\mu S_u^{cf} +
2 A_i \gamma_\nu {S'_u}^{cf} \gamma_\mu S_d^{be} +
\right. \nnb \\
&& \left. +
2 S_d^{be} \gamma_\nu A'_i \gamma_\mu S_u^{cf} +
2 A_i \tr (\gamma_\nu {S'_u}^{cf} \gamma_\mu S_d^{be}) +
\right. \nnb \\
&& \left. +
 S_d^{be} \tr (\gamma_\nu A'_i \gamma_\mu S_u^{cf}) +
\right. \nnb \\
&& \left. +
2 S_u^{cf} \gamma_\nu {S'_d}^{be} \gamma_\mu A_i +
2 S_u^{cf} \gamma_\nu A'_i \gamma_\mu S_d^{be} +
\right. \nnb \\
&& \left. +
2 S_d^{be} \gamma_\nu {S'_u}^{cf} \gamma_\mu A_i +
2 S_u^{cf} \tr (\gamma_\nu A'_i \gamma_\mu S_d^{be}) +
\right. \nnb \\
&& \left. +
S_d^{be} \tr(\gamma_\nu {S'_u}^{cf} \gamma_\mu A_i)
\right\} + \bar d^e A_i d^b \nnb \\
&&\left\{
2 S_u^{ad} \gamma_\nu A'_i \gamma_\mu S_u^{cf} +
2 S_u^{ad} \gamma_\nu {S'_u}^{cf} \gamma_\mu A_i +
\right. \nnb \\
&& \left. +
2 A_i  \gamma_\nu {S'_u}^{ad} \gamma_\mu S_u^{cf} +
2 S_u^{ad} \tr (\gamma_\nu {S'_u}^{cf} \gamma_\mu A_i) +
\right. \nnb \\
&& \left. +
A_i \tr (\gamma_\nu {S'_u}^{ad} \gamma_\mu S_u^{ad} )
\right\} \ve 0 \ra \label{deltap}
\eea
where $A_i = 1, \, \gamma_\alpha, \, \sigma_{\alpha \beta}/\sqrt{2}, \, i \gamma_\alpha \gamma_5,
\, \gamma_5$, a sum over $A_i$ implied, $S' \equiv CS^TC$, $A'_i = CA_i^TC$, with $T$ denoting the
transpose of the matrix, and $S_q$ is the full light quark propagator with both perturbative and
non-perturbative contributions:
\bea
S_q &=& \la 0 \ve {\cal T} \bar q(x) q(0) \ve 0 \ra \nnb \\
&=& \frac{i \not\!x}{2 \pi^2 x^4} - \frac{\la \bar q q \ra}{12} - \frac{x^2}{192} m_0^2\la \bar q q \ra
- \nnb \\ 
&-& i g_s \int_0^1 dv \left[ \frac{\not\!x}{16 \pi^2 x^2} G_{\mu \nu}(v x)  \sigma_{\mu \nu} - v x_\mu
G_{\mu \nu}(v x) \gamma_\nu \frac{i}{4 \pi^2 x^2} \right] \label{12}
\eea  
%%%%%%%%%%%%%%%%%%%%%%%%%

The $\Pi_{\mu \nu}'^\Delta$s in Eqs. (\ref{deltapp}) and (\ref{deltap}) describe diagrams in which the
photon interact with the quark lines perturbatively.  Their explicit expressions can be obtained from
the remaining terms by substituting all occurances of 
\bea
\bar q^a(x) A_i q^b {A_i}_{\alpha \beta} \rightarrow 2 \left( \int d^4 y  F_{\mu \nu} y_\nu S_q^{pert}
(x-y) \gamma_\mu S_q^{pert}(y)\right)^{ba}_{\alpha \beta} \label{pert}
\eea
where the Fock-Schwinger gauge is used, and $S_q^{pert}$ is the perturbative quark propogator, i.e. the
first term in Eq. (\ref{12}). 
%%%%%%%%%%%%%%%%%%%%%%%%%%%%%%%%%%%%%%%

The corresponding expressions for the correlation functions for 
the $\Delta^-$ and $\Delta^0$ baryons can be obtained from Eqs. (\ref{deltapp}) and (\ref{deltap}),
if one exchanges $u$-quarks by $d$-quarks and vice versa, respectively. 

In order to be able to calculate the QCD part of the sum rules, one needs to know the matrix elements
$\la \gamma (q) \ve \bar q A_i q \ve 0 \ra$.  Upto twist-4, the non-zero matrix elements given in terms
of the photon wave functions are [22-24]:
\bea
\la \gamma(q) \ve \bar q \gamma_\alpha \gamma_5 q \ve 0 \ra &=& \frac{f}{4} e_q  \epsilon_{\alpha \beta
\rho \sigma} \epsilon^\beta q^\rho x^\sigma \int_0^1 du e^{i u q x} \psi(u) \nnb \\
\la \gamma(q) \ve \bar q \sigma_{\alpha \beta} q \ve 0 \ra &=& i e_q  \la \bar q q \ra \int_0^1 du e^{i
u q x} \nnb \\
&\times& \{ (\epsilon_\alpha q_\beta - \epsilon_\beta q_\alpha) [\chi \phi(u) + x^2 [ g_1(u) - g_2(u) ]]
\nnb \\
&+& \left[ q x (\epsilon_\alpha x_\beta - \epsilon_\beta x_\alpha) + \epsilon x (x_\alpha q_\beta -
x_\beta
q_\alpha ) \right] g_2(u) \} \label{13}
\eea
where $\chi$ is the magnetic susceptibility of the quark condensate and $e_q$ is the quark charge.
The functions $\phi(u)$ and $\psi (u)$ are the leading twist-2 photon wave functions, while $g_1(u)$ and
$g_2 (u)$ are the twist-4 functions.

Note that, since we have assumed massless quarks, $m_u = m_d = 0$, and exact SU(2) flavor
symmetry, which implies $\la \bar u u \ra = \la \bar d d \ra$, the $u$ and $d$ quark propagators are
identical, $S_u = S_d$, whereas for the wave functions, the only difference is due to the different
charges of the two quarks.  The general expressions given by Eqs. (\ref{deltapp}) and (\ref{deltap}),
under exact $SU(2)$ flavor symmetry lead to the following results:
\bea
\Pi_{\mu \nu}^{\Delta^{++}} &=& - \frac{1}{2} e_u {\cal C} \nnb \\
\Pi_{\mu \nu}^{\Delta^{+}} &=& - \frac{1}{6} (2 e_u + e_d) {\cal C} \nnb \\
\Pi_{\mu \nu}^{\Delta^{0}} &=& - \frac{1}{6} (2 e_d + e_u) {\cal C} \nnb \\
\Pi_{\mu \nu}^{\Delta^{-}} &=& - \frac{1}{2} e_d {\cal C} \label{14}
\eea
where ${\cal C}$ is a factor independent of the quark charges. From Eq. (\ref{14}), the following exact
relations between theoretical parts of the correlator functions immediately follows:
\bea
\Pi_{\mu \nu}^{\Delta^+} &=& - \Pi_{\mu \nu}^{\Delta^-} = \frac{1}{2} \Pi_{\mu \nu}^{\Delta^{++}} \nnb
\\
\Pi_{\mu \nu}^{\Delta^0} &=& 0 \label{relations}
\eea
Hence, from now on, only the results for $\Delta^{++}$ will be reported and for the other $\Delta$'s,
the corresponding results can be obtained from the Eqs. (\ref{relations}).
Using Eqs. (\ref{12}) and (\ref{13}), from Eq. (\ref{deltapp}) and after some algebra 
and after performing Fourier transformation, for the coefficient of the structure \\ 
$g_{\mu \nu} \not\!p_1 \not\!\epsilon\not\!p_2$, we get:
\bea
\Pi &=&   e_u \int_0^1 du \left\{ \frac{f \psi (u)}{48 \pi^2} \left[ 4 \ln (-P^2) + \frac{\la
g^2G^2\ra}{12 P^4} \right] + \right. \nnb \\ 
&&\left. +
\frac{8}{3 P^4} \la \bar u u \ra^2 [g_1(u) - g_2(u)] + \frac{\chi
\phi(u) \la \bar u u \ra^2}{6P^2} \left( \frac{m_0^2}{P^2} + 4 \right) + \right. \nnb \\
&&+ \left.  \frac{2 \la \bar u u \ra^2}{3 P^4} - \frac{\la g^2 G^2 \ra}{768 \pi^4 P^2} - \frac{3 P^2
\ln (-P^2)}{64 \pi^4} \right\} \label{fourier}
\eea
where $P = p + q u$. In Eq. (\ref{fourier}), polynomials in $P^2$ are omitted since they do not
contribute after the Borel transformation.

As stated earlier, in order to obtain the sum rules, one equates the phenomenological and theoretical
expressions obtained within QCD. 
After performing the Borel transformation on the variables $p^2$ and $(p+q)^2$ in order to suppress the
contributions of the higher resonances and the continuum,  the following sum rules is obtained for the
magnetic moment of $\Delta^{++}$:
\bea
g_M &=&  \frac{3  e_u}{\lambda_{\Delta}^2} e^{\frac{M_{\Delta}^2}{M^2}}\left\{ \frac{f
\psi(u_0)}{12 \pi^2} \left[\frac{\la g^2 G^2 \ra}{48} -  M^4 f_1(\frac{s_0}{M^2}) \right] +
\right. \nnb \\
&+& \left. \frac{8}{3} \la \bar u u \ra^2 [g_1(u_0) - g_2(u_0)] + \right. \nnb \\
&+& \left. \frac{\chi \phi(u_0) \la \bar u u \ra^2}{6}
\left[m_0^2 -4 M^2 f_0(\frac{s_0}{M^2}) \right] \right. \nnb \\ 
&&+ \left.  \frac{2 \la \bar u u \ra^2}{3} +\frac{\la g^2 G^2 \ra M^2}{768
\pi^4}f_0(\frac{s_0}{M^2}) + \frac{3 M^6}{64 \pi^4}f_2(\frac{s_0}{M^2}) \right\} \label{sr}
\eea 
where the functions
\bea
f_n(x) = 1 - e^{-x} \sum_{k=0}^n \frac{x^k}{k!}
\eea
are used to subtract the contributions of the continuum. In Eq. (\ref{sr}), $s_0$ is the continuum
threshold, 
\bea
u_0 &=& \frac{M_1^2}{M_1^2 + M_2^2} \nnb \\
\frac{1}{M^2} &=& \frac{1}{M_1^2} + \frac{1}{M_2^2} \nnb
\eea
As we are dealing with just a single baryon, the Borel parameters $M_1^2$ and $M_2^2$ can be taken to be
equal, i.e. $M_1^2 = M_2^2$, from which it follows that $u_0 = 1/2$.

\section{Numerical Analysis}

From Eq. (\ref{sr}), one sees that, in  order to calculate the numerical value of the magnetic moment of
the $\Delta^{++}$, besides several numerical constants, one requires expressions for the photon wave
functions. It was shown in \cite{balitsky, braun} that the leading photon wave functions receive only
small corrections from the higher conformal spin, so they do not deviate much from the asymptotic form.  
Following \cite{braun,ali}, we shall use the following photon wave functions:
\bea
\phi(u) &=& 6 u \bar u \nnb \\
\psi(u) &=& 1 \nnb \\
g_1(u) &=& -\frac{1}{8} \bar u (3 - u) \nnb \\
g_2(u) &=& - \frac{1}{4} \bar u ^2 \nnb
\eea
where $\bar u = 1-u $.
The values of the other constants that are used in the calculation are: $f=0.028 \, GeV^2$, $\chi = - 
4.4 \, GeV^{-2}$ \cite{belyaev4} (in \cite{balitsky2}, $\chi$ is estimated to be $\chi = -3.3 \,
GeV^{-2}$), $\la g^2 G^2 \ra = 0.474 \, GeV^4$, $\la \bar u u \ra = - (0.243)^3 \, GeV^3$, $m_0^2 =
(0.8 \pm 0.2)\, GeV^2$ \cite{belyaev3}, $\lambda_\Delta = 0.038$ \cite{lee5}.

Having fixed the input parameters, our next task is to find a region of Borel parameter, $M^2$, where
dependence of the magnetic moments on $M^2$ and the continuum threshold $s_0$ is rather weak and at the
same time the higher dimension
operators, higher states and continuum contributions remain under control.
We demand that these contributions  are less then $35\%$.  Under this requirement, the working region 
for the Borel parameter, $M^2$, is found to be $1.1\, GeV^2 \le M^2 \le 1.4 \, GeV^2$.
Finally, in this range of the Borel parameter, the magnetic moment of $\Delta^{++}$ is
found to be $(4.55 \pm 0.03) ~ \mu_N$.  This prediction on the magnetic moment is obtained at $s_0=4.4
\, GeV^{2}$.  Choosing $s_0 = 3.8 \, GeV^2$ or $s_0=4.2 \, GeV^2$ changes the result at most by
$6\%$(see Fig. 1). The
calculated magnetic moment is  in good agreement with  the experimental result
 $(4.52 \pm 0.50 \pm 0.45) \, \mu_N$  \cite{wallace}. Our results on the magnetic moments for
$\Delta^{+}$, $\Delta^0$ and $\Delta^-$ are presented in Table 1.
For completeness, in this table, we have also presented the
predictions of  other approaches.  Comparing the values presented in Table 1, it is seen that our
predictions on magnetic moments is larger than the QCDSR predictions.

Finally, for the calculation of the magnetic moments of other members of the decuplet (which contain at
least one $s$-quark), the correction due to the strange quark mass should be taken into account.
Their calculations would be presented in a future work.

\newpage
\begin{table}[hu]
\caption{Comparisons of $\Delta$ baryon magnetic moments from
various calculations: this work (LCQSR),QCDSR \protect\cite{lee1}
lattice QCD (Latt)~\protect\cite{Derek92},
chiral perturbation theory ($\chi$PT)~\protect\cite{Butler94},
light-cone relativistic quark model
(RQM)~\protect\cite{Schlumpf93},
non-relativistic quark model (NQM)~\protect\cite{PDG92},
chiral quark-soliton model ($\chi$QSM)~\protect\cite{Kim97}, chiral
bag-model ($\chi$B) \cite{hong}.
All results are in units of nuclear magnetons.}
\label{comp}
\vspace{1cm}
\begin{tabular}{|l||c|c|c|c|}
\hline
Baryon	&	$\Delta^{++}$	&	$\Delta^+$	&	$\Delta^0$	&	$\Delta^-$ \\
\hline
\hline
Exp.	&	4.5$\pm$1.0	&			&	$\sim 0$	&			\\
\hline
LCQCD	&	4.4$\pm$0.8	&	2.2$\pm$0.4	&	0.00		&	-2.2$\pm$0.4	\\
\hline
QCDSR	&	4.13$\pm$1.30	&	2.07$\pm$0.65	&	0.00		& 	-2.07$\pm$0.65\\
\hline
Latt.	&	4.91$\pm$0.61	&	2.46$\pm$0.31	&	0.00		&	-2.46$\pm$0.31	\\
\hline
$\chi$PT&	4.0$\pm$0.4	&	2.1$\pm$0.2	&	-0.17$\pm$0.04	&	-2.25$\pm$0.25	\\
\hline
RQM	&	4.76		&	2.38		&	0.00		&	-2.38	\\
\hline
NQM	&	5.56		&	2.73		&	-0.09		&	-2.92	\\
\hline
$\chi$QSM&	4.73		&	2.19		&	-0.35		&	-2.90	\\
\hline
$\chi$B&	3.59		&	0.75		&	-2.09		&	-1.93	\\
\hline
\end{tabular}
\end{table}
\newpage
\begin{figure}
\vskip 1.5 cm
    \includegraphics{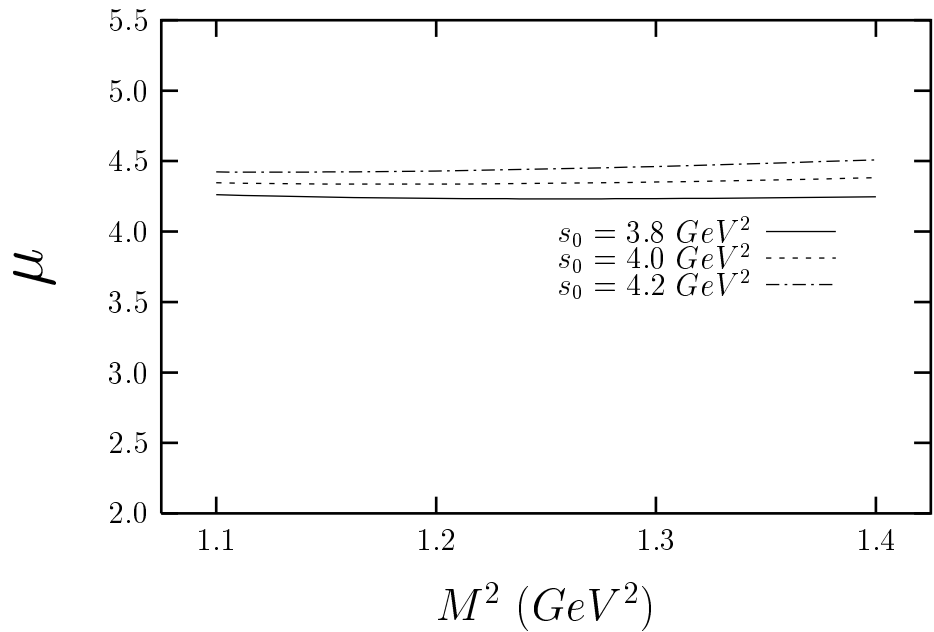}
\vskip 0.1cm
\caption{The dependence of the magnetic moment of $\Delta^{++}$ on the borel parameter $M^2$ (in
nuclear magneton units) for three different values of the continuum threshold, $s_0$.}
\end{figure}

\end{document}